\documentclass[twocolumn]{aastex62}

\usepackage{hyperref}
\usepackage{amsmath}
\usepackage{amssymb}
\usepackage{graphicx}
\usepackage[utf8]{inputenc}

\DeclareRobustCommand{\ion}[2]{\textup{#1\,\textsc{\lowercase{#2}}}}
\newcommand\kms{\ensuremath{\mbox{km}\,\mbox{s}^{-1}}}
\newcommand\Teff{\ensuremath{T_\mathrm{eff}}}
\newcommand\logg{\ensuremath{\log g}}

\newcommand\vt{\ensuremath{\xi_{t}}}

\newcommand\fei{\ion{Fe}{i}}
\newcommand\feii{\ion{Fe}{ii}}

\newcommand\sii{\ion{Si}{i}}
\newcommand\zni{\ion{Zn}{i}}
\newcommand\znii{\ion{Zn}{ii}}
\newcommand\he{HE\,1327$-$2326}

\graphicspath{{./}{figures/}}

\received{November, 2018}
\submitjournal{ApJ}
\shorttitle{Evidence for an aspherical supernova explosion of a Population\,III star}
\shortauthors{Ezzeddine et al.}

\begin{document}

\title{\Large {Evidence for an aspherical Population\,III supernova explosion inferred from the hyper metal-poor star HE\,1327$-$2326}\footnote{Based on observations made with the NASA/ESA Hubble Space Telescope, obtained at the Space Telescope Science Institute (STScI), which is operated by the Association of Universities for Research in Astronomy, Inc. (AURA) under NASA contract NAS 5-26555. These observations are associated with program GO-14151.}}

\correspondingauthor{Rana Ezzeddine}
\email{ranae@mit.edu}

\author[0000-0002-8504-8470]{Rana Ezzeddine}
\affil{Joint Institute for Nuclear Astrophysics, Center for the Evolution of the Elements, East Lansing, MI 48824, USA}
\affil{Department of Physics and Kavli Institute for Astrophysics and Space Research,\\ Massachusetts Institute of Technology, Cambridge, MA 02139, USA}
\affil{Department of Physics and Astronomy, Michigan State University, East Lansing, MI 48824, USA}

\author{Anna Frebel}
\affiliation{Department of Physics and Kavli Institute for Astrophysics and Space Research,\\ Massachusetts Institute of Technology, Cambridge, MA 02139, USA}
\affiliation{Joint Institute for Nuclear Astrophysics, Center for the Evolution of the Elements, East Lansing, MI 48824, USA}

\author{Ian U. Roederer}
\affiliation{Department of Astronomy, University of Michigan, 
 %1085 S. University Avenue, 
 Ann Arbor, MI 48109, USA}
\affiliation{Joint Institute for Nuclear Astrophysics, Center for the Evolution of the Elements, East Lansing, MI 48824, USA}

\author{Nozomu Tominaga}
\affiliation{Department of Physics, Faculty of Science and Engineering, Konan University, 
 %8-9-1 Okamoto, 
 Kobe, Hyogo 658-8501, Japan}
\affiliation{Kavli Institute for the Physics and Mathematics of the Universe, The University of Tokyo, 
 %5-1-5 Kashiwanoha, 
 Kashiwa, Chiba 277-8583, Japan}

\author{Jason Tumlinson}
\affiliation{Space Telescope Science Institute, Baltimore, MD 21218, USA}

\author{Miho Ishigaki}
\affiliation{Department of Astronomy, School of Science, The University of Tokyo, Bunkyo-ku, Tokyo 113-0033, Japan}

\author{Ken'ichi Nomoto}
\affiliation{Department of Astronomy, School of Science, The University of Tokyo, Bunkyo-ku, Tokyo 113-0033, Japan}
\affiliation{Kavli Institute for the Physics and Mathematics of the Universe, The University of Tokyo, 
 %5-1-5 Kashiwanoha, 
 Kashiwa, Chiba 277-8583, Japan}

\author{Vinicius M. Placco}
\affiliation{Department of Physics, University of Notre Dame, Notre Dame, IN 46556, USA}
\affiliation{Joint Institute for Nuclear Astrophysics, Center for the Evolution of the Elements, East Lansing, MI 48824, USA}

\author{Wako Aoki}
\affiliation{National Astronomical Observatory of Japan, Mitaka, Tokyo 181-8588, Japan}
\affiliation{Department of Astronomical Science, The Graduate University for Advanced Studies, Mitaka, Tokyo 181-8588, Japan}


\begin{abstract}
%Elements heavier than hydrogen and helium were first produced in the universe within the first stars. After a few million years, these presumably massive stars exploded as the first supernovae, ejecting the newly forged elements. Theoretical investigations have long indicated that such supernovae would explode in an asymmetric fashion, but insufficient observational evidence has prevented in-depth studies.
We present observational evidence that an aspherical supernova explosion could have occurred in the first stars in the early universe. 
Our results are based on the first determination of a Zn abundance in a \textit{Hubble Space Telescope}/Cosmic Origins Spectrograph high-resolution UV spectrum of a hyper metal-poor (HMP) star, \he,  with $\mbox{[Fe/H]}{\mathrm{(NLTE)}}=-5.2$. We determine $\mbox{[Zn/Fe]}=0.80 \pm 0.25$ from a UV \zni\ line at 2138\,{\AA}, detected at 3.4\,$\sigma$. Yields of a 25\,M$_{\odot}$  aspherical supernova model with artificially modified densities exploding with $E = 5 \times 10^{51}$\,ergs best match the entire abundance pattern of \he.
%Aspherical explosions in the early universe have already been suggested based on data from metal-poor stars, but firm conclusions were lacking, as these stars may have had multiple progenitor supernovae. 
Such high-entropy hypernova explosions are expected to produce bipolar outflows which could facilitate the external enrichment of small neighboring galaxies. This has already been predicted by theoretical studies of the earliest star forming minihalos. Such a scenario would have significant implications for the chemical enrichment across the early universe as HMP Carbon Enhanced Metal-Poor (CEMP) stars such as \he\ might have formed in such externally enriched environments.
\end{abstract}

\keywords{Galaxy: halo
$–$ nuclear reactions, nucleosynthesis, abundance $–$
stars: abundances $–$ stars: Population II $-$ stars: supernovae $-$ stars: individual (\he)}

\section{Introduction} \label{sec:intro}
%Characterizing the nature of the short-lived first stars and the properties and mechanisms of their supernova explosions (SNe) is indispensable to understand the subsequent chemical enrichment and  evolution that took place across the Galaxy and the early Universe. 
Our knowledge on the nature, formation and properties of the  Population\,III (Pop\,III) stars and their first supernova explosions (SNe) has been driven mainly  by theoretical work and cosmological simulations \citep{bromm2009,greif2010}. These studies have, for example, predicted a large range of masses for the first stars over the past decade,  from $<$1\,M$_{\odot}$ to $>$1000\,M$_{\odot}$ \citep{omukai2001,yoshida2006,stacy2012,hartwig2015}. 
 %Theoretical investigations have long indicated that such supernovae would explode in an asymmetric fashion \citep{maeda2003,tominaga2009,janka2013,grimmett2018}, but insufficient observational evidence has prevented in-depth studies. 
More generally, evidence on the first stars and SNe can be obtained from the chemical signatures of  surviving low-mass, ultra metal-poor (UMP) stars with $\mbox{[Fe/H]}$\footnote{Defined as $\mbox{[A/B]} = \log_{10}(N_{\mathrm{A}}/N_{\mathrm{B}})_{\mathrm{star}} - \log_{10}(N_{\mathrm{A}}/N_{\mathrm{B}})_{\odot}$, with $N_{\mathrm{A}}$ and $N_{\mathrm{B}}$ being the respective elements number densities.}$<-4.0$ \citep{Beers2005}. They likely formed from gas enriched by individual first SNe events whose chemical signature they have retained over billions of years \citep{Frebel2015,hartwig2018}. 
%Such evidence can be inferred by comparing the chemical abundance patterns of UMP stars to the theoretically predicted yields of their putative progenitor first star SNe.

%Thirty ultra metal-poor (UMP) stars with $\mbox{[Fe/H]}<-4.0$\footnote{Defined as $\mbox{[A/B]} = \log_{10}(N_{\mathrm{A}}/N_{\mathrm{B}})_{\mathrm{star}} - \log_{10}(N_{\mathrm{A}}/N_{\mathrm{B}})_{\odot}$, with $N_{\mathrm{A}}$ and $N_{\mathrm{B}}$ being the respective elements number densities.} \citep{Beers2005} have been identified over the past two decades \citep{frebel2018,Ezzeddine2017}, which opened a new and unique observational window into the early Universe. Such stars with extreme paucity in iron abundance, 
%Each of their stellar chemical signatures uniquely encode information on the properties of their first star progenitors, including their mass, explosion energy and mechanism, and details of the explosive production of elements \citep{nomoto2013}. 

Theoretical SNe nucleosynthesis yield calculations of  Pop\,III progenitors have long been invoked to explain the origins of the observed abundance patterns in UMP stars \citep{woosley1995,umeda2002,iwamoto2005,tominaga2007a,heger2010,grimmett2018}. 
Early spherical SNe models were not able to attain both the  $\alpha$-element and iron-peak (Mn, Co, Zn) abundance enhancements relative to iron \citep{cayrel2004}, irrespective of the explosion energies employed \citep{woosley1995}. Neither could they simultaneously produce the observed very large light element [C,N,O/Fe]\,$>1$ ratios found in most UMP stars. 

To better reproduce these ratios, a ``mixing and
fallback'' (MF) SNe model was developed \citep{umeda2002}, defined by three parameters to account for aspherical effects, namely (i) the initial mass cut which corresponds to the inner boundary of the mixing region ($M_{\mathrm{cut}}$), (ii) the outer boundary of the mixing region ($M_{\mathrm{mix}}$), and (iii) the
ejection factor ($f$) corresponding to the fraction of matter ejected from the mixed region.
 Two different mechanisms can be mimicked by
the MF process (as shown in Figures 12 (a) and (b) in \citealt{tominaga2007a}): (1) Faint, quasi-spherical MF SNe (with explosion energies $E\le 10^{51}$ erg) which experience
significant gravitational fallback of material onto the
nascent black hole. Before the fallback, Rayleigh-Taylor
instabilities induce mixing \citep{joggerst2010} of the
products of complete silicon burning (e.g., calcium, titanium
and iron) with material from the carbon-oxygen core. 
(2) Aspherical bipolar jet SNe which experience significant gravitational
fallback of material onto the nascent black hole along the
equatorial plane and ejection of material from the complete
Si burning layers along the jets \citep{tominaga2007b,tominaga2009}.
Both mechanisms result in the ejection of a relatively small (residual) amount of iron but a comparably large amount
of, e.g., carbon \citep{umeda2002}. This way,
the large ratios of [C/Fe], [N/Fe], and [O/Fe] in UMP stars could be produced. 

\begin{figure}
\begin{center}
\hspace*{-0.6cm}
\includegraphics[scale=0.16]{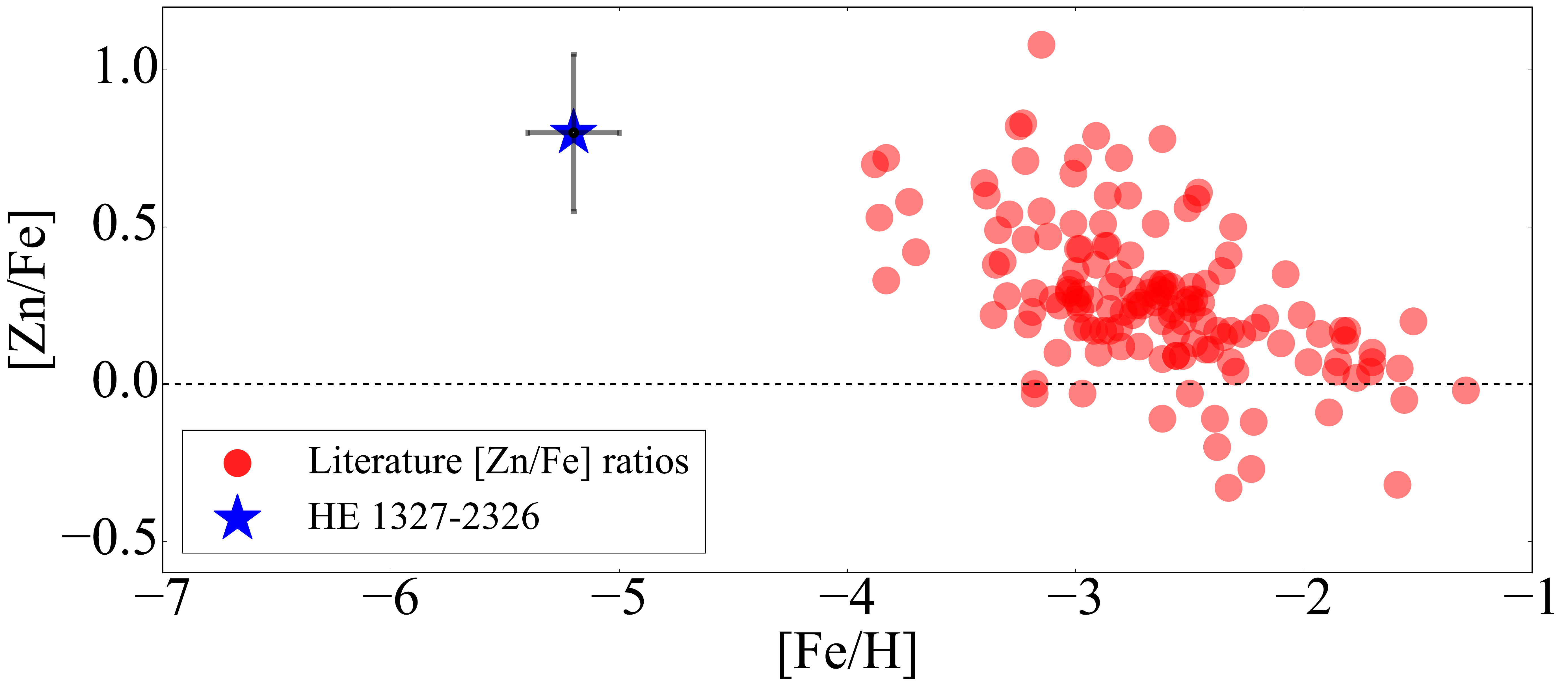}
\caption{\label{fig:zn_lit_he} \footnotesize [Zn/Fe] abundance ratios for metal-poor stars with $-4<\mbox{[Fe/H]}<-1$ from \citet{cayrel2004}, \citet{barklem2005}, \citet{hollek2011} and \citet{jacobson2015}. The $\mbox{[Zn/Fe]}=0.80\pm0.25$ obtained in this work for \he\ at $\mbox{[Fe/H]}=-5.20\pm0.20$ is shown by the blue star and black error bars. Abundance values are extracted from the JINAbase \citep{abohalima2018}.}
\end{center}
\end{figure}

The difference between the two mechanisms, however, appears in the abundance ratios among Fe-peak elements.
In the former quasi-spherical faint explosion model, large Zn
and other iron-peak elemental abundances cannot be
obtained \citep{tominaga2009,nomoto2013,grimmett2018} because a weak explosion energy is required to achieve the extensive fallback (necessary for producing a low iron abundance). On the other hand, in the aspherical bipolar jet explosion model, the high entropy environment along the jets enables large [Zn/Fe] ratios since more Zn is ejected.  Studies of extremely metal-poor stars with $-4.5 < \mbox{[Fe/H]} < -3$ have indeed shown enhanced $\mbox{[Zn/Fe]} > 0$ ratios
(Figure\,\ref{fig:zn_lit_he}; abundance data from \citealt{cayrel2004,barklem2005,hollek2011,jacobson2015}; abundances extracted from the JINAbase\footnote{https://jinabase.pythonanywhere.com}; \citealp{abohalima2018}), thus favoring the aspherical explosion model.

But no zinc measurements have thus far been possible for stars with [Fe/H]$<-4.5$ due to the intrinsic weakness of the strongest optical zinc triplet lines at 4680\,{\AA}, 4722\,{\AA} and $4810$\,{\AA}. However, another strong \zni\ line exists in the UV spectral range at $2138$\,{\AA}. Unfortunately, all stars with [Fe/H]$<-4.5$ are much too faint for successful UV observations---with the exception of \he, a Milky Way halo hyper metal-poor (HMP) star with $\mbox{[Fe/H](NLTE)}=-5.2$ \citep{Ezzeddine2018}. This star has been extensively studied based on optical spectra \citep{frebel2005,aoki2006,frebel2008,korn2009}. To continue to provide the most stringent observational constraints on the explosion mechanism, the details of the progenitor star and the origin of Zn in UMP stars, we obtained a \textit{Hubble Space Telescope}/Cosmic Origins Spectrograph (\textit{HST}/COS) high-resolution UV spectrum for \he\ in which we detected the absorption line of zinc at $2138$\,{\AA} for the first time. 

We describe the UV observations and data reduction in Section\,\ref{sec:obs} and the chemical abundance analysis in Section\,\ref{sec:method}. We use the [Zn/Fe] ratio in \he\ to constrain the SNe properties of its first star progenitor in Section\,\ref{sec:interp}, and present our interpretations and conclusions in Section\,\ref{sec:conclusion}.

%Portions of the spectra are shown in Figure~\ref{fig:spec}. We determined a zinc-to-iron ratio of $\mbox{[Zn/Fe]}=0.80 \pm 0.25$ (Zn line detected at 3.4\,$\sigma$-significance), which is the first one obtained for such an ancient second-generation star at $\mbox{[Fe/H]}<-4.5$. The Zn abundance in \he\ is enhanced relative to Fe, similar to what has been determined from optical Zn abundances in other ancient stars with $\mbox{[Fe/H]}>-4.0$. In Figure~\ref{fig:zn_lit_he}, we show the [Zn/Fe] ratio for \he\ in comparison with such old stars.

%We describe the UV observations and data reduction of the spectrum of \he\ in Sec.\,\ref{sec:obs}. In Sec.\,\ref{sec:method}, we describe the fundamental parameters used and the abundance analysis

\begin{figure*}[htp!]
\hspace{-2.5cm}
\includegraphics[scale=0.148]{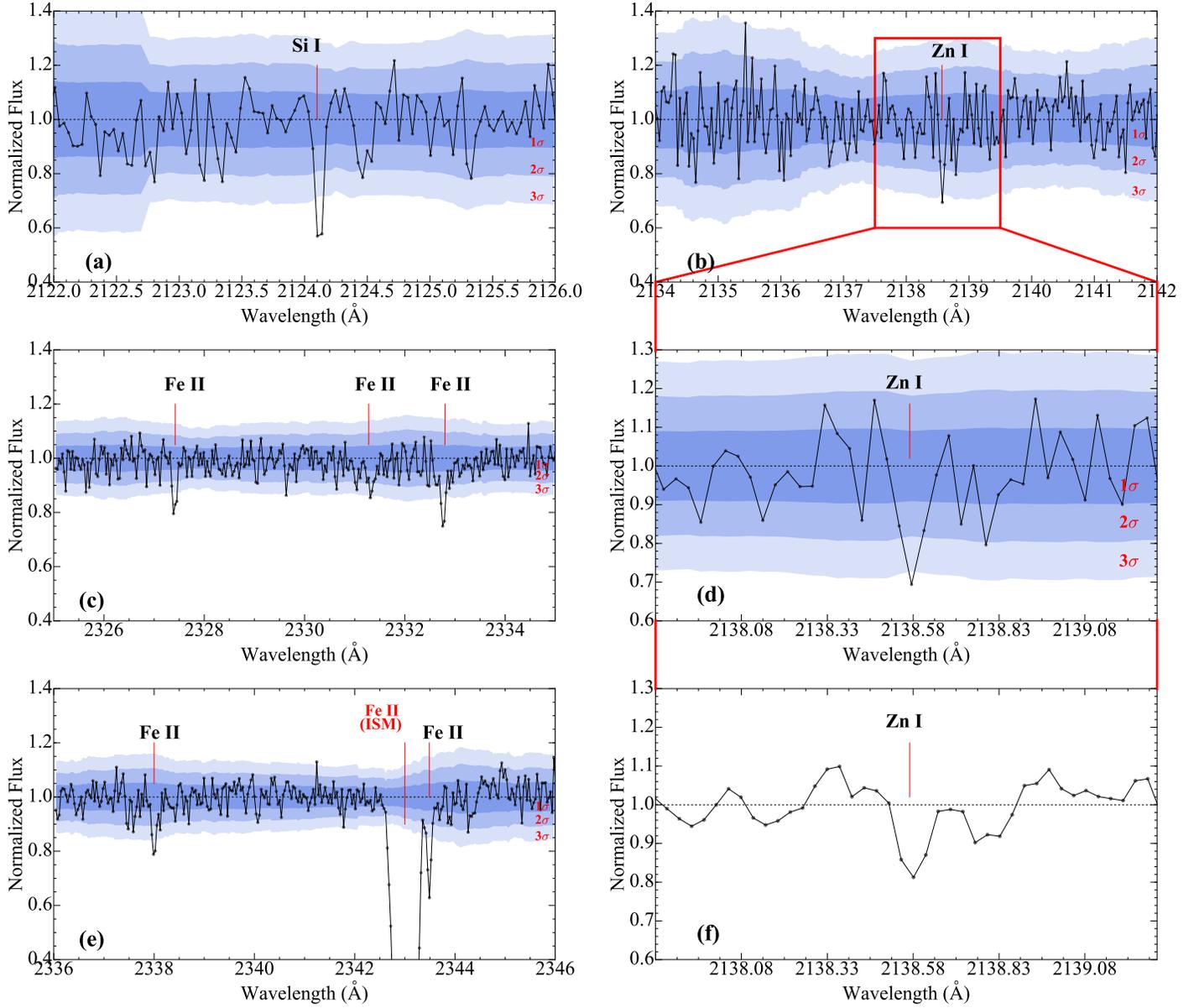}
\caption{Selected regions of the co-added UV spectrum of \he, where atomic lines of \sii, \zni\ and \feii\ were detected for the first time. The \feii\ interstellar medium (ISM) line at 2343\,{\AA} is also detected, as shown in panel (e). The detection significance levels of the lines are shown by the different shaded areas: $1\sigma$ (darkest blue), $2\sigma$ (light blue) and $3\sigma$ (lightest blue). For clarification purposes, panel (f) shows a smoothed part of the spectrum centered around the Zn line at 2138.57\,{\AA}, and covering the same spectral range as panel (d). The spectrum was smoothed using an average boxcar smoothing with a 2.0\,pixel width. For completeness, we also show the \sii\ and \feii\ spectral line regions, studied in details in \citet{Ezzeddine2018}.}
\label{fig:spec}
\end{figure*}

\section{Observations and data reduction} \label{sec:obs}
We obtained UV observations of \he\ with the Cosmic Origins Spectrograph (COS) on board the \textit{Hubble Space Telescope} (\textit{HST}) (Program ID: GO-14151), during May 07-15 2016.  The spectrum covers the three wavelength regions: $2118-2151$\,{\AA}, $2216-2249$\,{\AA} and $2315-2348$\,{\AA}.  The total integration time was 22\,hrs and 20\,minutes.
We performed a custom data reduction to optimize the signal-to-noise (S/N)
ratio of the final spectrum. By default, the COS extraction pipeline, CALCOS \citep{fox2015}, uses an extraction box height of 56 pixels in the cross dispersion (``y'') direction to extract the three stripes from the exposures. This procedure ensures that virtually all the source flux is captured, but at the cost of summing over pixels where the raw S/N is very low. Alternatively, we used an extraction box height of 9 pixels in the cross dispersion direction. This height collects somewhat less than the total source flux (at the 10\% level) and so can be used safely for line-to-continuum contrast measurements but not for cases where precise spectro-photometric measurements are necessary for the science. In our case, the vertical (cross-dispersion or ``y'') location of the 9-pixel extraction boxes was tuned individually for each stripe in each exposure using the B\_SPEC and HEIGHT parameters in the XTRACTAB reference file, and the edge-to-edge slope (SLOPE) in the diagonal boxes was retained at their default values. This is the procedure recommended by the STScI COS instrument team \citep{snyder2017}.  More details on the extraction, co-addition and normalization of the final UV spectrum can also be found in \citet{Ezzeddine2018}.

One Zn\,I line at 2138.57\,{\AA}, five \feii\ lines and one Si~\textsc{i} line, were detected in the final UV spectrum. Figure\,\ref{fig:spec} shows parts of these detections. For visual purposes, the bottom right panel (f) of Figure\,\ref{fig:spec} shows the Zn\,I line region smoothed with an average boxcar of a 2\,pixels width. The abundances of Fe and Si were determined and discussed further in \citet{Ezzeddine2018}. 

\section{Chemical abundance analysis}
\label{sec:method}
\subsection{Fundamental Stellar Parameters}\label{sec:stel_param}
In our abundance analysis of \he, we use an effective temperature of $\Teff=6180 \pm 80$\,K, derived from color-effective temperature relations using broad-band \textit{UBVRI} photometry \citep{frebel2005}, a surface gravity of $\logg=3.7 \pm 0.2$ and a microturbulent velocity of $\vt=1.7$\,\kms\, based on previous studies \citep{korn2009,Ezzeddine2018}. The \textit{Gaia} second data release DR2 \citep{gaia2018} produced 
$\Teff=5915\pm300$\,K and $\logg=3.40\pm0.3$ based on  photometric colors and parallax, respectively. We use both sets of stellar parameters to determine the [Zn/Fe] abundance ratio of \he. %Section\,\ref{iron_abnd} describes these results.
%We used these values to confirm our abundance results.
We employ a standard model atmosphere \citep{castelli2004} with input model metallicity of $\mbox{[Fe/H]}=-5.0$ and $\alpha$-enhancement $\mbox{[$\alpha$/Fe]}=0.4$  throughout. 

\subsection{$\mbox{[Zn/H]}$ and $\mbox{[Zn/Fe]}$ abundances in \he}\label{sec:zn_abund}
We measure an equivalent width of $43.4\pm12.7$\,m{\AA} for the \zni\ line at 2138.57\,{\AA}, by convolving the COS line-spread functions \citep{ghavamian2009} with Gaussian profiles following \citet{roederer2016}. We determine uncertainties in the equivalent width and Zn abundance measurements by altering the continuum placement and the full-width half maximum (FWHM) of the spectral line by $\pm 1\,\sigma$, and recording the corresponding changes. We assess a detection significance of 3.4$\sigma$ by dividing the equivalent width of the line by its uncertainty.
We then calculate the abundance of Zn using the equivalent width curve-of-growth method using the Local Thermodynamic Equilibrium (LTE) radiative transfer code \texttt{MOOG} updated to include proper scattering treatment \citep{sneden1973,sobeck2011}, and with custom analysis tools \citep{casey2014}. 
The abundances are determined relative to the solar values from \citet{asplund2009}. 

\begin{deluxetable}{l c r c r r r}
\tablewidth{0pt}
%\tabletypesize{\tiny}
\tablecaption{\label{tab:linelist} Chemical abundances of 13 elements determined from the UV and optical spectra of \he. Upper limits for Sc, Mn and Co are also included.}
\tablehead{
\colhead{El} & \colhead{N$_{\mathrm{nlines}}$} & \colhead{$\log \epsilon(X)$} & \colhead{$\sigma\log \epsilon(X)$} & \colhead{$\mbox{[X/H]}$} & \colhead{$\mbox{[X/Fe]}$}}
\startdata
C (CH)           & syn.  &  6.21\tablenotemark{a}  & 0.10&$-2.22$  &  3.49 \\
N (NH)\dotfill   & syn. & 6.10\tablenotemark{a} & 0.20 &$-1.73$ & 3.98 \\
O (OH) \dotfill  & syn. & 6.12\tablenotemark{a} & 0.20  & $-2.57$  & 3.14 \\
\ion{Na}{i}\dotfill  & 2 & 2.99  & 0.04 &$-3.25$ & 2.46 \\
\ion{Mg}{i}\dotfill  & 4&  3.54  & 0.02 &$-4.06$ & 1.65  \\
\ion{Al}{i}\dotfill  & 1&  1.90  & 0.03&$-4.55$ &  1.16  \\
 \ion{Si}{i}\tablenotemark{b}\dotfill   & 1 &  2.80 & 0.27 & $-4.71$ & 1.28 \\
\ion{Ca}{ii}\dotfill & 4  & 1.34 &0.15 & $-5.00$ & 0.71 \\
\ion{Ti}{ii}\dotfill  &15 & $-0.09$ & 0.17&$-5.04$ & 0.67  \\
\fei\tablenotemark{c}\dotfill & 10 & 1.79 & 0.15&$-5.71$ & \nodata\\
\fei\tablenotemark{d}\dotfill & 10  & 2.30 & 0.11 &$-5.20$ & \nodata\\
\feii \dotfill  & 4 &  1.51 & 0.26 &$-5.99$ & \nodata  \\
\ion{Ni}{i}\dotfill & 4 &  0.73 & 0.20 & $-5.49$ & 0.33\\
\ion{Zn}{i}\dotfill  & 1 &  $+0.16$ & 0.25 & $-4.40$ & $0.80$ \\
\ion{Sr}{ii}\dotfill & 2&  $-1.76$ &  0.06 & $-4.63$ & 1.08 \\
\ion{Sc}{ii}\dotfill & \nodata  &  $<-1.68$ & \nodata & $<-4.83$ & $<0.88$   \\
\ion{Mn}{i}\dotfill &  \nodata   &  $<0.53$ & \nodata & $<-4.90$  & $<0.81$ \\
\ion{Co}{i}\dotfill & \nodata  & $<0.58$ & \nodata & $<-4.41$ & $<1.30$ \\
\enddata
\tablenotetext{a}{3D values adopted from \citet{frebel2008}}
\tablenotetext{b}{Adopted from \citet{Ezzeddine2018}}
\tablenotetext{c}{LTE abundance from \citet{frebel2008}}
\tablenotetext{d}{NLTE abundance adopted from \citet{Ezzeddine2018}}
\end{deluxetable}

Using $\Teff=6180$\,K and $\logg=3.7$ \citep{frebel2008}, we determine a Zn abundance of $\mbox{[Zn/H]}=-4.40\pm0.25$ using the 1D, LTE framework. Conditions of line formation in metal-poor stars are known to deviate from the assumptions of LTE and 1D model atmospheres \citep{thevenin1999,mashonkina2011,bergemann2012,takeda2005}. Abundances from lines of minority species, such as neutral \zni\ and \fei, are often prone to larger non-local thermodynamic equilibrium (NLTE) effects in the atmospheres of low metallicity stars than from line of the corresponding dominant ionized species. 
Investigations of possible NLTE effects for abundances of commonly used optical \zni\ lines over a range of stellar parameters showed that NLTE abundance corrections, defined by the differences between the NLTE and LTE abundances using the same observed equivalent widths, are actually not significant and $\leq 0.2$\,dex \citep{takeda2005}. Even though the departures from LTE have not been calculated explicitly for any UV \zni\ lines, it could be shown by \citet{roederer2018} that ionization equilibrium in LTE can be reached in metal-poor stars with $-3.0<\mbox{[Fe/H]}<-1.0$ (within 0.1\,dex) using optical (e.g., \zni\ $\lambda4810$) and UV (e.g., \znii\ $\lambda2062$) lines. Departures from LTE for Zn were indeed insignificant, even at $\mbox{[Fe/H]}=-3.0$. Following these results, we assume that our LTE Zn abundance, as derived from the UV line, would be compatible with values obtained from any optical lines, if ever detected.

For Fe, however, strong NLTE effects have been well documented (e.g., \citealt{asplund2005}, and references therein). We thus adopt the 1D, NLTE Fe abundance of $\mbox{[Fe/H]}=-5.20$ in \he, inferred from 10 \fei\ optical lines as determined in \citet{Ezzeddine2018}. This Fe abundance for \he\ is based on detailed investigations of possible 3D and NLTE effects on both \fei\ and \feii\ lines in the optical and UV. We then obtained a zinc to iron abundance ratio of $\mbox{[Zn/Fe]}=0.80\pm0.25$. 
For completeness, we note that any potential NLTE correction  to the \zni\ abundance (should they exist) would be positive, and would thus only further increase the [Zn/Fe] ratio \citep{takeda2005}.

We also investigates the [Zn/Fe] abundance ratio using the \textit{Gaia} DR2 stellar parameters which yield slightly lower 1D, LTE Zn and 1D, NLTE Fe abundances of $\mbox{[Zn/H]}=-4.70$ and $\mbox{[Fe/H]}=-5.40$, respectively. This 1D, NLTE abundance of Fe was determined following the same setup used in \citet{Ezzeddine2017}.
Compared to our other value, the $\mbox{[Zn/Fe]}$ decreases only slightly, to $0.70\pm0.25$. This shows that the $\mbox{[Zn/Fe]}$ ratio in \he\ is robustly enhanced, irrespective the choice of stellar parameters.

To establish the overall [X/Fe] abundance pattern of \he, we adopt $\mbox{[Zn/Fe]}=0.80\pm0.25$, in addition to the UV \sii\ abundance from \citet{Ezzeddine2018} as well as the optical abundances for 11 elements and upper limits for five others (Sc, V, Cr, Mn and Co)  from \citet{frebel2008}. All elemental abundances used are listed in Table\,\ref{tab:linelist}. We use this pattern to compare to the theoretical SNe nucleosynthesis yields of various models.

\section{Constraining the Pop\,III star explosion properties}\label{sec:interp}
\subsection{Comparison to faint mixing and fallback quasi-spherical SNe models}

Previous studies have commonly compared abundance patterns of individual UMP stars, such as that of \he, to the yields of faint mixing and fallback quasi-spherical Pop\,III SNe \citep{umeda2002,iwamoto2005,frebel2005,christlieb2004,keller2014,bessell2015,placco2016b,nordlander2017}. 
As discussed in Section\,\ref{sec:intro}, such type of spherical explosion underproduces  iron peak elements (such as Co, Cr, or Zn) \citep{tominaga2009,grimmett2018,hirai2018,tsujimoto2018}. Due to a lack of iron peak abundances, the comparisons of the data with these model results usually remained unconstrained in the that region. With available Zn abundances, however, new constraints can  be obtained. 
%We also examined the chemical ejecta prior to the supernova explosion of rotating and non-rotating progenitors.

\begin{figure}
\begin{center}
\hspace*{-1.0cm}
\includegraphics[scale=0.37]{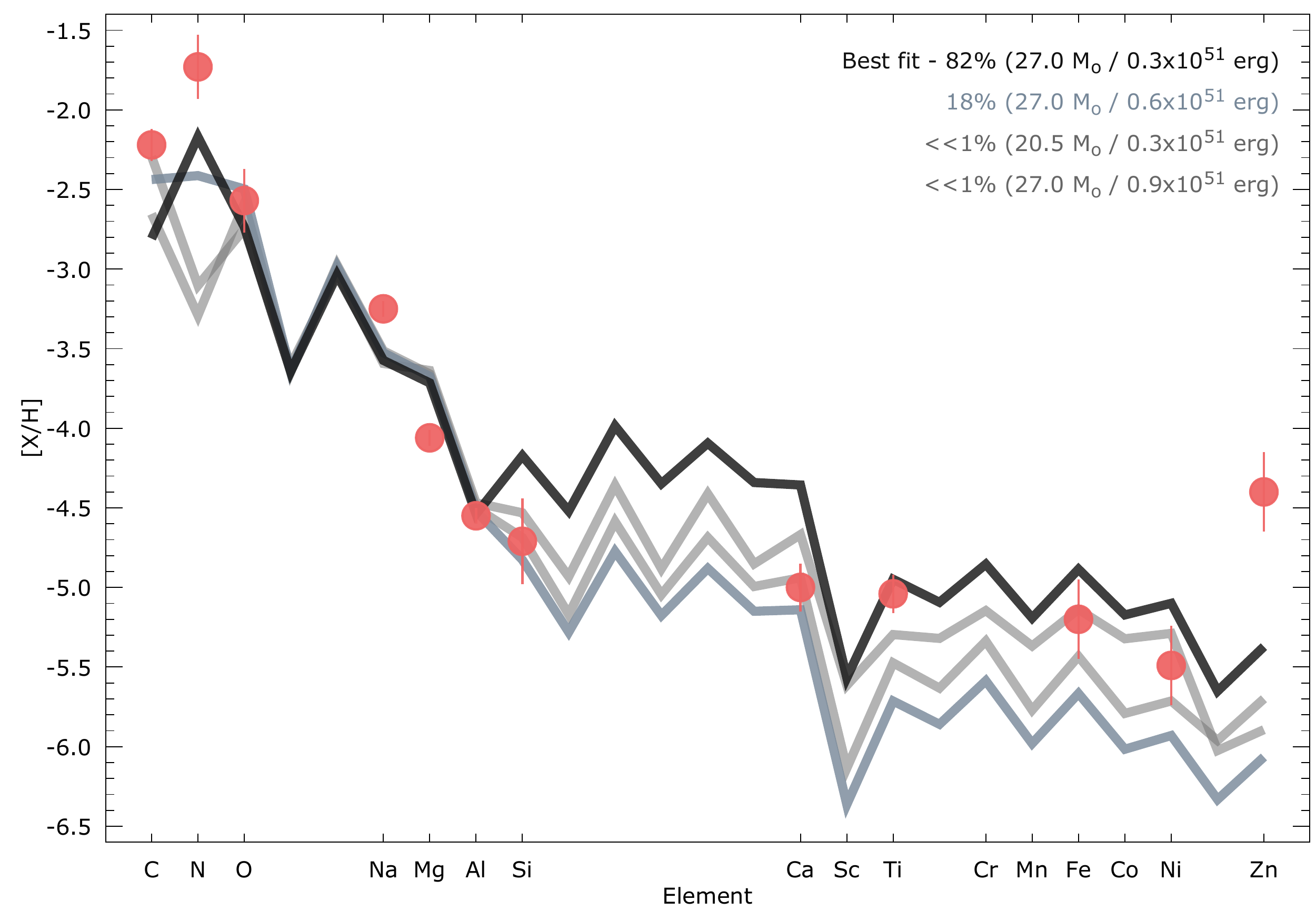}
\caption{Statistical results for comparisons of a large grid of (16,800) mixing and fallback (quasi-spherical) SNe nucleosynthesis yields \citep{heger2010} with different progenitor masses and explosion energies, to 10,000 re-sampled chemical abundance patterns of \he. }
\end{center}
\label{starfit}
\end{figure}

We perform statistical fitting tests following \citet{placco2016b}, to compare the determined abundance pattern of \he\ (including Zn) to 16,800 theoretical yields\footnote{Models from https://2sn.org/starfit/ \citep{heger2010}} of quasi-spherical mixing and fallback SNe models, computed with different explosion parameters (e.g., progenitor mass, explosion energies, mixing parameters). To account for uncertainties in our fitting results, we simulate 10,000 different abundances for each observed element used in the fit, which follow a normal distribution centered around the measured abundance (taken as the mean). We then randomly sample the abundance distributions for all elements and generate 10,000 artificial abundance patterns for \he, to compare the yields. Our results, shown in Figure\,\ref{starfit}, demonstrate that indeed none of the mixing-fallback supernova models can match the Zn abundance (relative to Fe) in \he. Overall, 82\% of the simulated abundance patterns of \he\ can be fit best (although not satisfactorily) with a 27\,M$_{\odot}$ progenitor exploding with $0.3\times10^{51}$\,erg. 18\% of the simulated abundance patterns are also reproduced by a 27\,M$_{\odot}$ progenitor, but with a slightly higher explosion energy of $0.6\times10^{51}$\,erg. The remaining few abundance patterns ($<<1\%$) are fit best with 20.5\,M$_{\odot}$ and 27\,M$_{\odot}$ progenitors with $0.3\times10^{51}$\,ergs and $0.9\times10^{51}$\,ergs, respectively.
%talk about the residuals of the fits and what exactly the right hand panel means!

Building on these results, we can statistically rule out faint quasi-spherical SNe as the source of metals in \he, as no superposition of yields would produce enough Zn relative to the lighter elements. Enrichment by multiple events can also more broadly be eliminated following the theoretical mono-enrichment classification scheme suggested by \citet{hartwig2018}, in which stars with $-6.0<\mbox{[Fe/H]}<-4.0$ and [Mg/C]$<-1.0$ have a strong likelihood of being enriched by a single SNe event. This applies well to \he\ with $\mbox{[Fe/H]}=-5.2$ and $\mbox{[Mg/C]}=-1.84$. Overall, this shows that the progenitor of \he, and possibly those of other UMP stars with similar abundance patterns, did not explode (quasi-)spherically. 

\begin{figure*}[htp!]
\begin{center}
\vspace*{-1cm}
\includegraphics[scale=0.29]{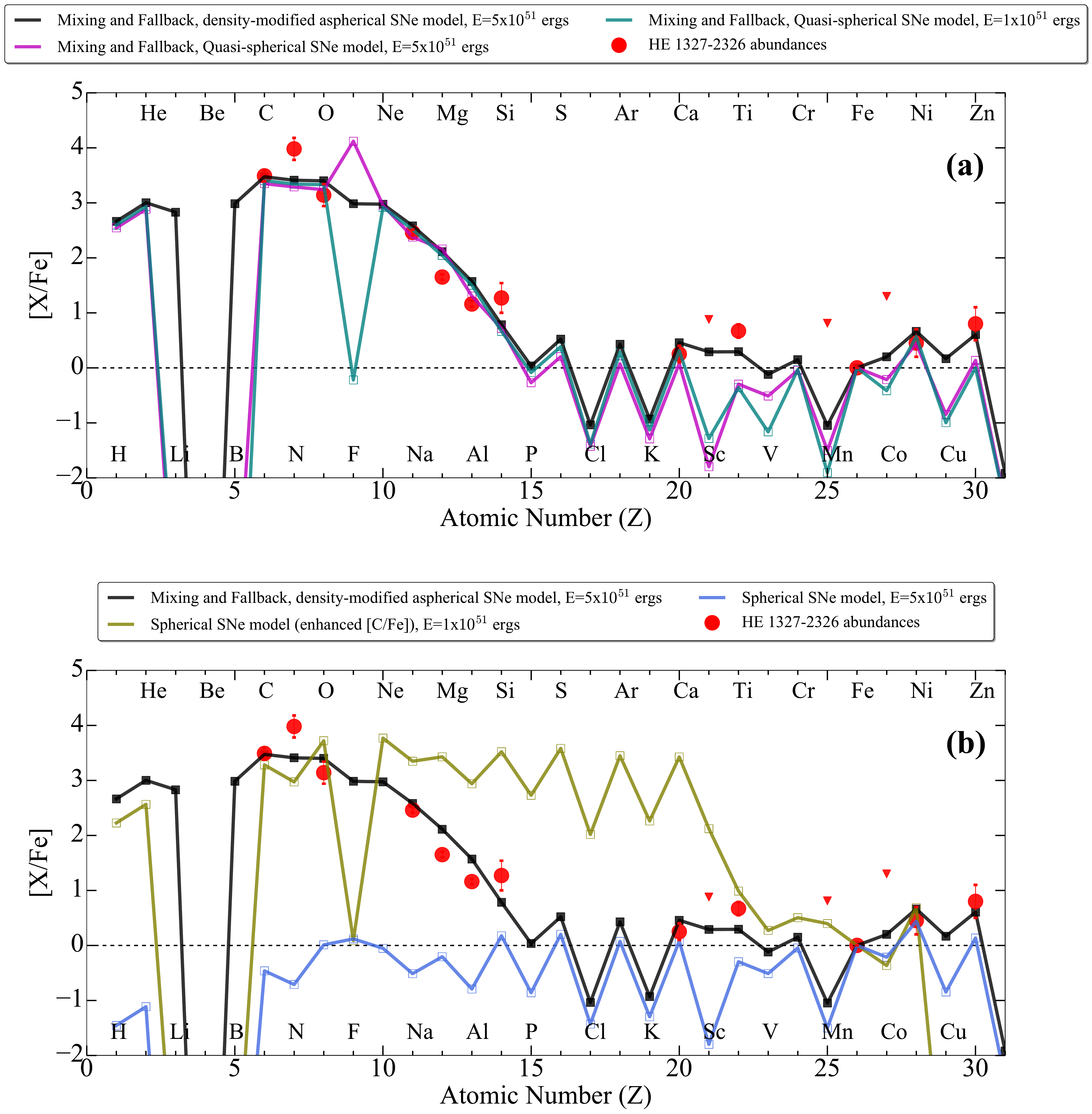}
\caption{Best-fit yields of an artificially
density-modified ``mixing and fallback" mimicking an aspherical SNe explosion model with bipolar-outflows of a first star progenitor with $25\,M_{\odot}$ and $E_{51} = 5 \times 10^{51}$\,ergs explosion energy (black solid line). This is obtained from fitting various models and associated parameters to the measured abundance pattern of HE~1327$-$2326 (red symbols). Triangles indicate upper limits. Panel (a) also shows yields for two $25\,M_{\odot}$ quasi-spherical mixing and fallback explosion models with explosion energies of $E = 1\times 10^{51}$\,ergs (solid teal line) and $E = 5\times 10^{51}$\,ergs (solid magenta line), respectively. Panel (b) shows yields for two spherical SNe explosion models with $E = 1\times 10^{51}$\,ergs and enhanced [C/Fe] ratio (solid gold line), and $E = 5\times 10^{51}$\,ergs (solid blue line), respectively. Only the artificially density modified, aspherically mimicked model with bipolar jets is able to reproduce the largely enhanced [Zn/Fe] abundance ratio determined in \he.} 
\end{center}
\label{SN_model_bestfit_aspherical}
\end{figure*}

\subsection{Comparison to Aspherical SNe models}\label{sec:asphericalSNe}
%Aspherical SNe with bipolar-outflows offer means to successfully reproduce the observed abundances in \he. 
We compare the abundance pattern of \he\ to the yields of artificially density modified mixing and fallback models with different masses and explosion energies, thus mimicking aspherical SNe with bipolar outflows. Specific details on the explosion models and yield calculations are described in \citet{tominaga2007a}.
We find that the yields of a $25\,M_{\odot}$ first star progenitor exploding with an aspherical SNe, with a high explosion energy of $E = 5 \times 10^{51}\,$ergs best matches the entire abundance pattern of \he, including zinc. We show this result by the black solid line in Figure~\ref{SN_model_bestfit_aspherical}. Our best fit model has an initial mass cut of $M_{\mathrm{cut}}$=$1.64\,M_{\odot}$, outer boundary mixing region mass of $M_{\mathrm{mix}}=5.65\,M_{\odot}$ and an ejection factor of $f=0.0002$.
To increase specifically the zinc yield of a mixing and fallback SNe model, the explosion mechanism was modified such that the matter density is artificially reduced, to mimic a high entropy explosion environment, which is required for increased explosive nucleosynthesis \citep{maeda2003,tominaga2007b,tominaga2007a,grimmett2018}. 
In this modified framework, elements formed in the deepest layers during the explosion, such as zinc, are (at least partially) mixed upward and then ejected. This occurs alongside the release of only small amounts of iron (most iron falls back onto the nascent black hole) and large amounts of, e.g., carbon made in the layers further out. The density artificially modified mixing and fallback model thus parameterizes what would physically be an aspherical SNe with bipolar outflows or jets (See panel (b) of Figure\,12 in \citealt{tominaga2007a}),
with the ejection factor $f$ in the 1D model being equivalent to the fraction of the solid angle of the region where the ejected mass elements in the Si-burning layer are located in the 2D model (i.e., $f=1$ in the spherical model).
The necessary high-entropy environment would occur along the rotation axis.
\citet{tominaga2009} showed that the abundance patterns of the yields of the 2D aspherical bipolar explosion is reproduced by the density modified mixing and fallback model.
%There are some elements showing differences - Sc, Ti, V, Cr, Co, and Zn. These elements are enhanced in the high-entropy environment in the bipolar explosion. 
%These thermodynamic features can be mimicked by artificially reducing the matter density \citep{tominaga2009}.}
%\textbf{This interpretation as an aspherical
%jet-like SNe has been further confirmed by more comprehensive 2-dimensional calculations of supernova explosions in previous studies \citep{maeda2003,tominaga2007b,tominaga2009}, which reproduce the abundance patterns of the metal-poor stars as with the mixing-fallback model.
Unfortunately, more
realistic 3D explosion models of metal-free stars are not
available yet to further investigate this behaviour.

%This interpretation as an aspherical jet-like SNe has been further confirmed by more comprehensive 2-dimensional calculations of supernova explosions in previous studies \citep{maeda2003,tominaga2007,tominaga2009}, which qualitatively show the same behavior.  Unfortunately, more realistic 3D explosion models of metal-free stars are not available yet to further investigate this behaviour. 

%\begin{figure}
%\begin{center}
%\hspace*{-0.8cm}
%\includegraphics[scale=0.3]{HE1327_50s3_edited.pdf}
%\caption{\label{cartoon}\footnotesize Illustration of an aspherical SNe exploding with bipolar outflow jets at $t = 50$\,s after the explosion, where the velocity decreases to $4\times10^{4}$\,\kms \citep{tominaga2009}. Colors represent densities of dominant ejected elements: Fe and Zn (green) and C and O (magenta).}
%\end{center}
%\end{figure}

We also explore other geometries using the same progenitor mass of $25\,M_{\odot}$ including two quasi-spherical mixing and fallback SNe models with low ($E=1\times10^{51}$\,ergs; teal solid line in panel (a) of Figure\,\ref{SN_model_bestfit_aspherical}) and high ($E=5\times10^{51}$\,ergs; magenta solid line in panel (a) of Figure\,\ref{SN_model_bestfit_aspherical}) explosion energies. Two spherical SNe models with and without enhanced [C/Fe] ratios were also explored corresponding to, respectively, the gold and blue solid lines in panel (b) of Figure\,\ref{SN_model_bestfit_aspherical}. 
The explosion parameters $M_{\mathrm{cut}}$,  $M_{\mathrm{mix}}$ and $f$ used for each of these models are listed in Table\,\ref{tab:models}.
Both quasi-spherical mixing and fallback models with $E=1\times10^{51}$\,ergs and $E=5\times10^{51}$\,ergs are not able to produce the [Zn/Fe]$>0$ ratio. Simultaneously, they also under-reproduce the [Sc/Fe] and [Ti/Fe] ratios as compared to the aspherical model. 
The explosive nucleosynthesis in the density modified, high-entropy aspherical model enhances the $\alpha$-rich freezeout, thus enhancing the [Sc/Fe] and [Ti/Fe] ratios usually determined in extremely metal-poor (EMP) stars with $\mbox{[Fe/H]}<-3$ \citep{tominaga2007a}. Given that the determined [Ti/Fe]=0.67 ratio in \he\ is enhanced, an aspherical explosion model is thus clearly favoured.

We note that spherical models could principally also produce a high entropy environment, and hence, a solar [Zn/Fe] ratio if the explosion energy were large enough, e.g., $E=5\times10^{51}$\,ergs (see panel (b) of Figure\,\ref{SN_model_bestfit_aspherical}) \citep{nomoto2013}. However, such an energetic explosion would be highly inconsistent with the significant fallback required to produce the low Fe abundances observed in UMP stars, including \he. This explains why the spherical models are unlikely contenders for the explosion mechanisms of \he's progenitor, and perhaps even for other first stars. 
%
%The spherical model yields with no [C/Fe] enhancement, and $E=5\times10^{51}$\,ergs under-produces the full abundance pattern of \he relative to the aspherical model (panel (b) of Figure\,\ref{SN_model_bestfit_aspherical}), except for the iron-peak element ratios [Co/Fe] and [Ni/Fe].
%Such an energetic explosion in spherical symmetry is highly inconsistent with the significant fallback required to produce the low Fe abundances observed in UMP stars. On the other hand, the spherical model with enhanced [C/Fe] ratio, is better able to reproduce the CNO elements, whereas highly over-producing the $\alpha$ elements as compared to the aspherical model.
%
%\citet{frohlich2006b,frohlich2006a}

\begin{deluxetable*}{l c c c c c c}
\tablewidth{0pt}
%\tabletypesize{\tiny}
\tablecaption{\label{tab:models} Parameters of the different SNe model yields used in Figure\,\ref{SN_model_bestfit_aspherical}.}
\tablehead{
\colhead{Model} & \colhead{Mass} & \colhead{$E$} & \colhead{$M_{\mathrm{cut}}$} & \colhead{$M_{\mathrm{mix}}$} & \colhead{$f$}\\
\colhead{} & \colhead{[$M_{\odot}$]} & \colhead{[$10^{51}$\,ergs]} & \colhead{[$M_{\odot}$]} & \colhead{[$M_{\odot}$]} & \colhead{}}
\startdata
Mixing and Fallback, density-modified aspherical & 25  &  5  & 1.64 & 5.65 & 0.0002 \\
Mixing and Fallback, quasi-spherical, high energy & 25  &  5  & 1.59 & 5.69 & 0.0001\\
Mixing and Fallback, quasi-spherical, low energy  & 25  &  1  & 1.61 & 5.65 & {  }{ }0.00015 \\
Spherical, enhanced [C/Fe] & 25  &  1 & 2.24 & \nodata & 1 \\
Spherical  & 25  &  5 & 1.59 & \nodata & 1 \\
\enddata
\end{deluxetable*}

%Without the density modification, quasi-spherical mixing-fallback models deliver qualitatively similar results except for a zinc enhancement. 
%Previous work
%, before the [Zn/Fe] ratio was determined and taken into account for \he, 
%has favoured a faint $E = 0.3 \times 10^{51}\,$ergs quasi-spherical mixing-fallback explosion model \citep{umeda2002} of a $25\,M_{\odot}$ progenitor \citep{iwamoto2005,placco2016b}, before the [Zn/Fe] ratio was determined and taken into account.
%
%
Overall, the explosion of the massive Pop\,III star progenitor of \he\ is thus more energetic than previously thought (i.e., a hypernova), in line with more recent studies, such as \cite{grimmett2018}.
%thus implying that  the first stars were perhaps significant producers of UV radiation, and contributed to the reionization of the universe\cite{bowman2018}. 
Interestingly, previous work has suggested that only faint mixing and fallback first SNe would seed carbon-enhanced metal-poor (CEMP) stars such as \he, since their host minihalos would be disrupted by any larger energy input \citep{cooke2014}. Our results provide an alternative explanation, since high energy aspherical SNe explosions could also produce high [C/Fe] ratios.

%\textcolor{red}{Here!Talk about Tominaga and Grimmet suggestions at such geometries using 2D and 1D approximated methods which could not have been applied to second-generation stars as no observations were available!}
%This broadly agrees with the results \cite{grimmett2018} that approximated aspherical (1D) supernova explosions with bipolar outflows can reproduce the abundances of old stars with $\mbox{[Fe/H]}>-4.0$ and enhancements in Zn and C relative to Fe. 

\subsection{Evidence for aspherical explosions from rotation}\label{sec:rot}

It has been proposed that aspherical SNe are driven by (fast) rotating progenitors with possibly strong magnetic fields \citep{maeda2003,meynet2006,ekstrom2008}.
This is in line with theoretical predictions that the first stars were heavily rotating \citep{meynet2006,stacy2011,tsujimoto2018}. Indeed, the progenitors of \he\ and those of other second generation stars have already been shown to be fast rotating based on their large relative nitrogen abundance ratios, e.g.,  $\mbox{[N/Fe]}=3.98$ in \he, $\mbox{[N/Fe]}=2.57$ in HE~0107$-$5240 \citep{christlieb2004}, and $\mbox{[N/Fe]}=3.46$ in SD~1313$-$0019 \citep{frebel2015b}.

Stellar surface nitrogen abundances become enhanced due to internal mixing caused by rotation \citep{meynet2006,choplin2017}, rather than being produced during their explosion (see Figure\,\ref{SN_model_bestfit_aspherical}).
Additionally, \citet{maeder2015} and \citet{choplin2018} have proposed that a fast rotating Pop\,III star progenitor could be responsible for the enhanced [Sr/Fe]=1.08  determined for \he\, and other EMP stars with $\mbox{[Sr/Fe]}>0.5$. 
%They showed that the ejecta[xxxx] of a  fast-rotating,
They showed that fast rotation in a low-metallicity, massive star can  set off a strong mixing between the H- and He-burning zones which  triggers the synthesis of Sr and other light neutron-capture elements made in the s-process.

Introducing rotation of Pop\,III stars into the standard neutrino-driven paradigm of core-collapse supernovae has independently been shown to lead to bipolar (jet-like) SNe \citep{fryer2000,burrows2004}. Finally, rotation has also been implemented into early universe and first stars cosmological simulations \citep{greif2010,stacy2011} to investigate the ab-initio formation of the first stars.
Moreover, \citet{tsujimoto2018} recently predicted through their galactic chemical evolution calculations that magneto-rotational driven explosions could indeed be the dominant source of enhanced [Zn/Fe] abundance ratios in the Milky Way satellite dwarf galaxy stars with $-4<\mbox{[Fe/H]}<-1$.

\subsection{$\nu p$-process in SNe: Is it able to produce the enhanced [Zn/Fe] in \he?}

\citet{frohlich2006b} and \citet{pruet2006} proposed that increasing the entropy by increasing the electron mole fractions $Y_{e}$ in a faint quasi-spherical SNe model to values $>0.5$ would mimic a $\nu p$-process in the neutrino driven winds emerging from the proto-neutron star of the Pop\,III progenitor. This could enhance the production of [Sr/Fe] as well as the iron peak elements [Sc,Co,Zn/Fe] ratios, as is observed in UMP stars.

\citet{frohlich2006a}, however, showed that such a scenario only leads to the production of Zn of up to  [Zn/Fe]\,$=0$ which remains inconsistent with observational results of [Zn/Fe]$\gtrsim0.5$, as determined for EMP stars and \he. \citet{tominaga2007a} confirmed these results by introducing neutrino-transport into their mixing and fallback SNe models of Pop\,III stars by artificially increasing Y$_{e}$ in the complete Si burning region where Zn is produced. They showed that increasing $Y_{e}$ to $>0.5$ could indeed enhance the production of $\mbox{[Sc/Fe]}>0.3$ as is observed in EMP stars but could not simultaneously produce the enhanced $\mbox{[Zn/Fe]}>0$ as well as $\mbox{[Co/Fe]}>0$.

%In general, an increase in entropy is required to produce the enhanced [Zn/Fe] ratios determined in EMP stars. This can be achieved by either 
%
%(i) introducing low-density modifications in the progenitor star during the SNe, i.e., mimicking an aspherical SNe with bipolar jets \citet{tominaga2007a,grimmett2018}, or 
%
%(ii) by increasing the electron mole fractions $Y_{e}$ to values $>0.5$, i.e., mimicking neutrino driven winds with the $\nu p$-process (see recent review by \citealt{thielemann2018review}).
%[why not start the paragraph with this? ]

%to 0.5. 

%They showed that their results are consistent with those obtained by  \citet{frohlich2006a}. 

This suggests that the $\nu$p process is not responsible for the enhanced Zn production in \he\ and other EMP stars. Instead, the entire abundance pattern of \he\ can be produced by rotation-driven, high energy ($E=\times10^{-51}$\,ergs) aspherical SNe with bipolar-jets, as described in Sections\,\ref{sec:asphericalSNe} and \ref{sec:rot}. Nevertheless, the $\nu p$ process could contribute to an increased [Sr/Fe] ratio as has been determined for \he\ and other UMP stars.

%Zn/Fe]=0, i.e. solar values, can be made in regular core-collapse SNe Frohlich et al. 2006a, Curtis et al. 2018), but the upturn (with a sizable scatter) below [Fe/H]=-2 has to be due to hypernovae and or a certain class of magneto-rotational supernovae (Tsujimoto \& Nishimura2018). (Thielemann)}

\section{Interpretation and Conclusions: Evidence for external chemical enrichment}\label{sec:conclusion}

Stellar abundance trends of various elements show increasing amounts of scatter with decreasing iron abundance. The fraction of stars with unusual abundance patterns is the highest at the lowest [Fe/H], with $\sim$80\% for $\mbox{[Fe/H]}<-4.0$, including \he. This has previously been interpreted as a result of inhomogeneous metal mixing or unusual SNe at the earliest times \citep{Frebel2015}. On the contrary, the vast majority of metal-poor stars with $\mbox{[Fe/H]}>-4.0$ show patterns generally in line with a formation from well mixed gas \citep{cayrel2004,yong2013b}. It is thus possible that these different types of abundance patterns are the result of different chemical enrichment channels that could have operated in the early universe. If one channel was common and the other rare, it might  explain the extreme rarity of stars with $\mbox{[Fe/H]}<-4.5$ which remain to have a poorly understood origin scenario. This idea is principally supported by the fact that models for metallicity ([Fe/H]) distribution function cannot reproduce the existence of the low-[Fe/H] tail with $\mbox{[Fe/H]}<-4.0$ \citep{yong2013b} simultaneously with the body of data on metal-poor stars with higher [Fe/H] abundances. 

These challenges might be explained by the existence of aspherical Pop\,III SNe, as described in this work. The ejecta structure of a bipolar explosion is different than that of a quasi-spherical explosion, in that the high energy bipolar explosion produces high velocity ejecta along the polar axis as well as along the equatorial plane. These velocities can reach up to $4\times10^{4}$\,\kms\ and $8\times10^3$\,\kms after 50\,s of the explosion, respectively \citep{tominaga2009}. This is higher than what can be maximally produced by faint mixing-fallback explosions ($4\times10^3$\,\kms) \citep{iwamoto2005}. 
 A high-velocity ejecta could facilitate carrying the SNe yields out of the parent host minihalo to enrich a neighboring minihalo. 
 %\he\ and the other UMP stars with enhanced [Zn/Fe] ratios would have formed in these externally enriched systems. We note that this would be true regardless of their carbon abundances as aspherical supernovae produce high carbon abundances.  Indeed, these stars make up at best only a few percent of known old stars, in line with predictions for the occurrence rate of such a rare channel \citep{hartwig2018}. 
 Theoretical studies have already explored this external enrichment channel across minihalos \citep{smith2015,jeon2017,hartwig2018}. Assuming the gas in the neighboring system to be primordial, the abundance ratios of the supernova yields should largely remain preserved, such as the high [C/Fe] ratio in combination with the low iron abundances, as well as enhanced [$\alpha$/Fe] and [Zn/Fe] values.  Such a channel could have important implications for our understanding of the first chemical enrichment events and how these are preserved in second-generation UMP stars, especially those highly enhanced in carbon, such as the CEMP star \he.

 An external enrichment scenario is principally supported by the outer halo nature of \he\ and similar metal-poor stars \citep{tissera2014,batagglia2017}. As such, they are likely accreted from some smaller now disrupted system, such as a primordial minihalo. The accretion origin of \he\ is based on its proper motion values from the latest \textit{Gaia} DR2 kinematic parameters for proper motions ($\mu_{\mathrm{RA}}=-52.52\pm0.04$\,mas\,yr$^{-1}$ and $\mu_{\mathrm{DEC}}=45.50\pm0.04$\,mas\,yr$^{-1}$) and parallax ($\pi=0.887\pm0.023$\,mas) used with a Galactic dynamical model \citep{carollo2014,frebel2018}.
 
 Stars such as \he\ might then have actually formed in these externally enriched neighboring primordial systems long before those were accreted into the Milky Way, as part of its own hierarchical growth. Future cosmological simulations will be able to provide additional insight into the nature of this channel, or provide alternative explanation for the formation sites of second-generation CEMP stars with the lowest [Fe/H] abundances.

%rather than wiped out by dynamical processes such as homogenization during recollapse following the energy injection by the first supernova(e)\cite{greif2010}.

%This presumably rare birth environment might explain the chemical signatures of these second-generation stars which are qualitatively significantly different (e.g. low [Fe/H] and high [C/Fe]) from those of other old stars at higher [Fe/H] values.  
%Future investigations are bound to show if indeed stars such as \he\ and others with similar abundance patterns, could thus have formed in such an externally enriched system.

%\section{Conclusions}

\acknowledgements 
We thank the anonymous referee for the careful reading of our manuscript and the insightful comments and suggestions. We also thank Arthur Choplin, Volker Bromm, John Wise, Britton Smith and Myoungwon Jeon for useful discussion. 
R.E., A.F., and I.U.R. acknowledge generous support for program HST-GO-14151 provided by a grant from STScI, which is operated by AURA, under NASA contract NAS~5-26555.
R.E., A.F., I.U.R. and V.P. acknowledge support from JINA-CEE, funded in part by the National Science Foundation under Grant No. PHY-1430152. A.F. is supported by NSF-CAREER grant AST-1255160. I.U.R. acknowledges additional support from NSF grants AST-1613536 and AST-1815403.\\

\software{MOOG~\citep{sneden1973,sobeck2011}, MULTI~\citep{Carlsson1986,carlsson1992},
MARCS~\citep{gustafsson1975,gustafsson2008}}
 
\bibliography{ref}

\end{document}